# Part2: Ultra-short pulse laser patterning of very thin indium tin oxide on glass substrates

C. McDonnell[1], D.Milne[2], H Chan., D. Rostohar[1], and G.M. O'Connor[1],

[1]*School of Physics, National Centre for Laser Applications, National University of Ireland Galway, Ireland*

[2]*M-Solv UK Ltd., Oxonian Park, Langford Locks, Kidlington, Oxford, OX5 1FP, United Kingdom*

**Corresponding Author:** Cormac McDonnell, National Centre for Laser Applications, National University of Ireland, Galway, Ireland

E-mail: c.mcdonnell3@nuigalway.ie Tel: +353 91 493 595

**Keywords:** ITO, thin films, laser, ablation, threshold fluence, femtosecond, transparent conductive layers, two temperature model

**Key highlights:**
1. Threshold fluences for ultrashort femtosecond and picosecond processes are found to be an order of magnitude less than those obtained for nanosecond laser pulses.
2. The ablative mechanism is found to be primarily by non-thermal ejection of the ITO grains driven by the hot electron blast force which is generated in the film.
3. Selective laser ablation of the film is possible, however selective removal is highly dependent on the applied laser fluence.
4. Line removal of the ITO film with overlapping pulses is shown to be not influenced on the percentage overlap of the laser pulses

*Abstract*

We investigate selective patterning of ultra-thin 20 nm Indium Tin Oxide (ITO) thin films on glass substrates, using 343, 515, & 1030 nm femtosecond (fs), and 1030 nm picoseconds (ps) laser pulses. An ablative removal mechanism is observed for all wavelengths at both femtosecond and picoseconds time-scales. The absorbed threshold fluence values were determined to be 12.5 mJcm$^{-2}$ at 343 nm, 9.68 mJcm$^{-2}$ at 515 nm, and 7.50 mJcm$^{-2}$ at 1030 nm for femtosecond and 9.14 mJcm$^{-2}$ at 1030 nm for picosecond laser exposure. Surface analysis of ablated craters using atomic force microscopy confirms that the selective removal of the film from the glass substrate is dependent on the applied fluence. Film removal is shown to be primarily through ultrafast lattice deformation generated by an electron blast force. The laser absorption and heating process was simulated using a two temperature model (TTM).

The predicted surface temperatures confirm that film removal below 1 Jcm-2 to be predominately by a non-thermal mechanism.

# 1. Introduction

The application of laser technology to selectively pattern ultra-thin transparent conducting oxides (TCO's) is important to the large area electronics value chain. Selective laser patterning has a number of advantages over traditional lithography-based processes; the process enables the development of clean, reconfigurable, large area patterning of diverse materials. The emergence of robust, high average power, ultra-short laser sources offer alternatives over the more established nanosecond laser sources. This study investigates such ultra-short selective patterning processes on very thin indium thin oxide (ITO) which facilitates a direct comparison with that previously presented for nanosecond lasers[1]. Better understanding of the interaction between an ultra- short laser pulse and the very thin ITO material on a thin glass substrate is needed to fully realise the potential of these future high repetition rate ultra-short laser sources. ITO as an n-type degenerate semiconductor, is likely to remain an important thin film TCO in the display and photovoltaic device sector due to its high transparency in the visible region, high electrical conductivity, mechanical durability and quality of its thin film deposition [1].

The indium tin oxide ($In_2O_3:Sn$), film structure is typically polycrystalline, with a grain size that is dependent on the deposition conditions, the thickness of the film, and the dopant concentration (%$SnO_2$). The crystal structure of indium oxide is bixbyite, with 16 $In_2O_3$ molecules in each unit cell [2]. The $Sn$ substitutional dopant replaces $In$, donating free electrons which increases the electronic conductivity [3]. The electronic structure of ITO is defined by the $Sn$ doping; doping results in a Fermi level $E_f$ located above the conduction band edge $E_c$. The values of $E_f, E_c$ and $E_g$, the band gap, all vary with doping concentration and deposition conditions. The direct band gap is typically around 3.8 - 4 eV, with a large density of free electrons at the bottom of the conduction band [4], typically in the range of $10^{20}/10^{21} cm^{-3}$ [5].

Effective selective removal of the ITO thin films is required for high volume precision manufacture. Laser processes with excellent electronic isolation properties, with minimal damage to the glass substrates, and in the cases of display devices, minimal line visibility are all required. The characteristics of laser patterned ITO films have been studied across a range of wavelengths in the nanosecond pulsed time regime [6-9], however, typically these processes have only focussed on selective removal of relatively thick ITO films with thicknesses of the order of 100 – 500 nm. More recent research has focussed on the use of ultra-short laser sources. For instance picosecond laser sources at 266, 355, and 532 nm were used to selectively remove thick (120 nm) ITO films [10]. 266 nm was identified as the optimal laser source for clean film removal using highly overlapped laser pulses. The film removal process was identified to be by spallation at 355 nm, with evidence of thermal damage at 532 nm. The removal of 150 nm thick ITO films from soda lime glass was also investigated using an IR 150 fs duration pulse, across a range of repetition rates and scan speeds [11]. Picosecond laser pulse have also shown promise for selective removal of 200 nm thick ITO films using 355, 532, and 1064 nm pulses [12]. Currently, no studies have been undertaken on the interaction of very thin ITO films with ultrashort laser pulses. As well as this, the exact removal mechanism of the film from the substrate is still currently under investigation.

The objective of the current study is to investigate the wavelength dependence of the single shot threshold fluence for ultra-short pulse selective removal of very thin ITO films. The laser threshold fluence, measured in Jcm$^{-2}$, is an important parameter in defining the laser material interaction; it is defined here as the fluence required initiating visible damage to the ultra-thin film. The dependence of the ablative processes on the applied fluence is also reported. A simple two-temperature, finite element, model was also developed to show the electron and lattice response to ultra-short laser pulses. The paper identifies the key differences between ultra-short and nanosecond selective patterning of very thin ITO films, as well as proposing the removal mechanism of the film, thereby informing future applications for such processes.

## 2. Materials and Methods

### 2.1 Materials

Polycrystalline ITO thin film samples were deposited on soda lime glass (10X-FS Corning), by DC sputtering (Aimcore, Taiwan). The very thin ITO film had a sheet resistances of < 80 Ω/□, a resistivity of $2.8 \times 10^4 \Omega cm$, and an estimated carrier concentration of $\sim 3.2 \times 10^{20} cm^{-3}$ The film thickness was verified experimentally to be 20 nm, with an average grain size in the range of 20 - 30 nm. The reflectivity and transmission of the ITO film was determined using a UV-VIS spectrometer, with a wavelength range of 200 - 1100 nm. An estimate for the absorption coefficient was determined from the Beer-Lambert law,

$$\alpha(\lambda) = \left(-\frac{1}{d}\ln\left[T(\lambda)/\{(1-R'(\lambda))(1-R(\lambda))\}\right]\right) \quad (1)$$

where $T(\lambda)$ is the measured transmission, $R(\lambda)$ is the measured reflectivity at ITO/air interface, $R'(\lambda)$ is the estimated reflectivity at ITO/glass interface, and $d$ is the film thickness. This equation takes into account both the reflected component from the ITO/air interface, and the ITO/glass interface.

The reflectivity between ITO on glass interface is estimated using the refractive indices of glass [13] and ITO [14], with the reflectivity given by the equation $R'(\lambda) = [(n_{ITO} - n_g)/(n_{ITO} + n_g)]^2$. This results in ITO/glass interface reflectivities of 0.045, 0.030, 0.011, and 0.005 for wavelengths of 266, 355, 532 and 1064 nm, respectively. The accuracy of transmission values are ± 0.006. While the actual absorption coefficient may be different at higher fluences due to non-linear absorption, the applied fluences in this study are considered low (< 2 Jcm$^{-2}$).

### 2.2 Measurement methodology

Laser patterning experiments were performed at a wavelength of 1030 nm using a 10 picosecond pulse duration (Trumpf Trumicro5050) and at three wavelengths of 343, 515 and 1030 nm using a 500 femtosecond pulse duration (Amplitude Systems S-pulse). The experimental setup is shown in Figure 1, consisting of the laser source, external attenuators, beam steering mirrors (Scanlabs Hurryscan) and focussing optics (Sill Optics, large area telecentric lens, f = 103 mm) was used in all cases. The laser spot radius at the focus was verified experimentally using Liu's method [15].

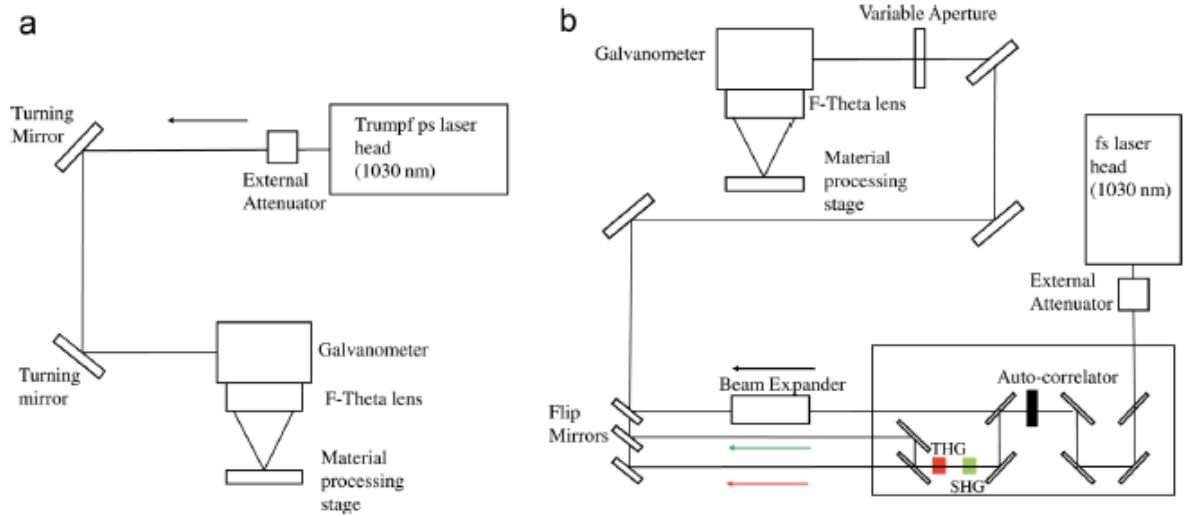

figure 1 Schematic diagram of the experimental configuration for femtosecond and picosecond pulse durations..

The laser pulses were incident on samples mounted on a horizontal plane at the focal point in such a way that the sample holder did not interfere with the transmitted beam. The details of the laser sources and configurations are reported in Table 1.

Table 1 Parameters for the femtosecond and picosecond laser wavelengths used for ITO thin film ablation

| Wavelength (nm) | Pulse Duration (fs) | Repetition Rate (kHz) | Raw Beam Output (mm) | Focal Length (mm) | NA | Spot radius (µm) |
|---|---|---|---|---|---|---|
| 343 | 500 | 1 | 2 | 103 | 0.004 | 10.2 |
| 515 | 500 | 1 | 2 | 103 | 0.004 | 24.5 |
| 1030 | 500 | 1 | 2 | 103 | 0.004 | 27.8 |
| 1030 | 10,000 | 10 | 3 | 103 | 0.024 | 23.1 |

Laser ablation experiments in the femtosecond time regime were performed at three wavelengths using a chirped pulse, regeneratively amplified, laser (Amplitude Systems S-pulse). For a specific pulse repetition rate, the overlap of the laser spots was controlled by adjusting the scanning speed of the galvanometer. Low (1 kHz) repetition rates were used to create spatially resolved single craters. The laser power at the workpiece was controlled using an external attenuator. Laser ablation experiments in the picosecond time regime were performed using a thin disk amplified laser system from Trumpf (TruMicro 550). Pulses were generated at the fundamental wavelength of 1030 nm, with an approximate pulse duration of 10 ps.

The average crater diameters were measured using an optical microscope (Olympus stm-mjs2). Atomic Force Microscopy (AFM) (Agilent 5500) was used to determine the film grain structure and the morphology of the laser patterned features. A silicon nitride cantilever (Nanosensors PPP-Contr, resonance frequency 13 $kHz$, tip radius $< 7\ nm$, force constant $\sim 0.2\ N/m$) was used in contact mode with a scan speed of 1 line/s.

Experiments were performed using Gaussian laser sources. Plotting $D^2$ versus the natural log of the fluence and extrapolating $D^2 = 0$ provides a measure of the applied threshold fluence [15]. As the film thickness used in this case is much smaller than the typical absorption depth of the ITO film, only a fraction of the applied laser fluence is absorbed in the film, therefore, a more useful measure of the threshold fluence can be found in terms of the absorbed threshold fluence, $\phi_{th}^{abs}(\lambda)$, which shows the threshold fluence in terms of the actual absorbed laser energy component,

$$\phi_{th}^{abs}(\lambda) \approx \phi_{th}(\lambda)\big(1 - R(\lambda)\big)(d_{film}\alpha(\lambda)), \qquad (2)$$

where $\phi_{th}(\lambda)$ is the applied threshold fluence, $R(\lambda)$ is the film reflectivity, and $d_{film}\alpha(\lambda)$ is the fraction of energy absorbed in the film through the absorption coefficient, assuming no reflection at the film glass interface. This method assumes that the film reflectivity and absorption coefficient are constant during the laser pulse. In experimental cases the high intensity laser pulse can result in non-linear absorptions in the film, however in this study the applied fluences are quite low, which we assume results in minimal non linear absorptions.

## 2.3 Computational laser heating simulation

In order to estimate the temperature in the ultra thin ITO film and glass substrate using ultrashort laser pulses, a two temperature model (TTM) was used [16]. While the TTM is typically used for metals, we propose that it is also relevant for ITO as it has a high free carrier electronic density, which dominates its optical absorption properties. The TTM defines the electronic and lattice as two separate subsystems. The temperature of the electrons $T_e$ and $T_l$ lattice is given by

$$C_e(T_e)\frac{\partial T_e}{\partial t} = \nabla(k_e(T_e)\nabla T_e) - G(T_e - T_l) + Q \qquad (3)$$

$$C_l(T_l)\frac{\partial T_l}{\partial x} = \nabla(k_l \nabla T_l) + G(T_e - T_l) \qquad (4)$$

where $C_e$, $C_l$ are the heat capacities of the electrons and lattice (JK$^{-1}$m$^{-3}$), $k_e$, $k_l$ are the heat conductivities of the electrons and lattice (Wm-1K-1) and $G$ is the electron phonon coupling factor (Wm$^{-3}$K$^{-1}$) and $Q$ is the laser heat source (Wm$^{-3}$). The laser source term is given by

$$Q(r,z,t) = \frac{2\phi_0 \alpha(\lambda)}{\sqrt{\pi/ln2}\tau_p}(1 - R(\lambda))exp\left\{-\frac{2r^2}{\omega_0^2} - (4ln2)\left(\frac{t}{\tau_p}\right)^2 - (\alpha(\lambda)z)\right\}, \qquad (5)$$

Where $\phi_0$ is the applied peak fluence, $R(\lambda)$ is the film reflectivity, $\tau_p$ is the full width half maximum (FWHM) pulse duration, $r$ is the radial coordinate, $\omega_0$ is the $1/e^2$ spot radius, $\alpha(\lambda)$ is the material absorption coefficient and $z$ is the depth component into the material. The model assumes optical energy is first absorbed by the electronic subsystem, and then transferred to the lattice through electron phonon coupling. An estimation of the electronic heat capacity and its dependence on the electron temperature, is given as $C_e = C_0 T_e$ [17] where $C_0$ is given as

$$C_0 = \frac{\pi^2 n_e k_B}{2T_F} \qquad (6)$$

Where $n_e$ is the density of electrons, $k_B$ is Boltzmann's constant, and $T_F$ is the Fermi temperature given as $T_F = E_F/k_B$. The thermal conductivity of the electrons is given as

$$k_e(T_e) = v^2 C_e(T_e)\tau_e/3 \tag{7}$$

where $v^2$ is the mean squared velocity of the electrons contributing to the electronic thermal conductivity. This component can be approximated as $v^2 = v_F^2$, where $v_F$ is the Fermi velocity, where $v_F = \sqrt{2E_F/m_e^*}$. $\tau_e$ is the electron scattering time, including contributions from electron-electron and electron-phonon scattering terms, $1/\tau_e = 1/\tau_{e-e} + 1/\tau_{e-ph}$. The electron phonon coupling factor G is given as

$$G = \frac{\pi^2 m_e^* c_s^2 n_e}{6\tau T_e} \tag{8}$$

where $m_e^*$ is the effective mass of the electron, $c_s$ is the speed of sound given as $c_s = \sqrt{\beta/\rho}$, where $\beta$ is the bulk modulus and $\rho$ is the material density, and $\tau$ is the electron relaxation time. Table 2 shows the parameters used for the ITO film in the two temperature model. The absorption coefficient of the film was calculated using (2). Due to the lack of data on the bulk modulus of ITO, the value of used in this in the simulation was simplified to that used for gold films in similar TTM models [18].

Table 2 Optical and material constants used for ITO thin film TTM model at λ = 1030 nm.

| Parameter | Symbol | Value | Unit |
|---|---|---|---|
| Number density of electrons | $n_e$ | $10^{21}$ | $cm^{-3}$ |
| Boltzmann's constant | $k_B$ | $1.38 \times 10^{-23}$ | $m^2\ kg\ s^{-2}\ K^{-1}$ |
| Lattice heat capacity | $C_l$ | $2.4 \times 10^6$ [19] | $Jm^{-3}\ K^{-1}$ |
| Fermi energy | $E_F$ | $3.3 / 5.29 \times 10^{-19}$ [2] | eV/ J |
| Mass of the electron | $m_e$ | $9.109 \times 10^{-31}$ | kg |
| Electronic effective mass in ITO | $m_e^*$ | $0.35 m_e$ [2] | kg |
| Electron scattering time | $\tau_e$ | 6 [20] | fs |
| Speed of Sound | $c_s$ | $5.56 \times 10^3$ | m/s |
| Pulse duration | $\tau_p$ | 500 (10) | fs (ps) |
| Abs. coefficient (1030 nm) | $\alpha(\lambda)$ | $1.2 \times 10^6\ m^{-1}$ | $m^{-1}$ |
| Reflectivity (1030 nm) | $R(\lambda)$ | 0.05 | - |

The purpose of the simulation is to guide the interpretation of results. There are approximations and assumptions used in the two temperature thin film simulation, including its dependence on the various static parameters (Table 2), its non-quantum treatment of the various material parameters, and the assumption that there is little or no re-distribution of energy via ballistic electrons, and the omission of nanoparticle formation and emission. Despite this, the simulation gives a reasonable approximation to the laser process; simulating values for $T_e$ and $T_l$ allows the determination of overall temperature evolution across a range of applied fluences, which can then be compared to the surface topography observed experimentally, which in turn provides insights into the film removal process

## 3. Results

### 3.1 Focussed threshold fluence of ITO thin films at ultrashort pulse durations

The threshold fluence for ITO thin film was measured at femtosecond and picosecond pulse durations, at wavelengths of 343, 515 and 1030 nm. Figure 2 shows the variation in the crater diameter with the applied fluence, for the ultrashort pulse durations used.

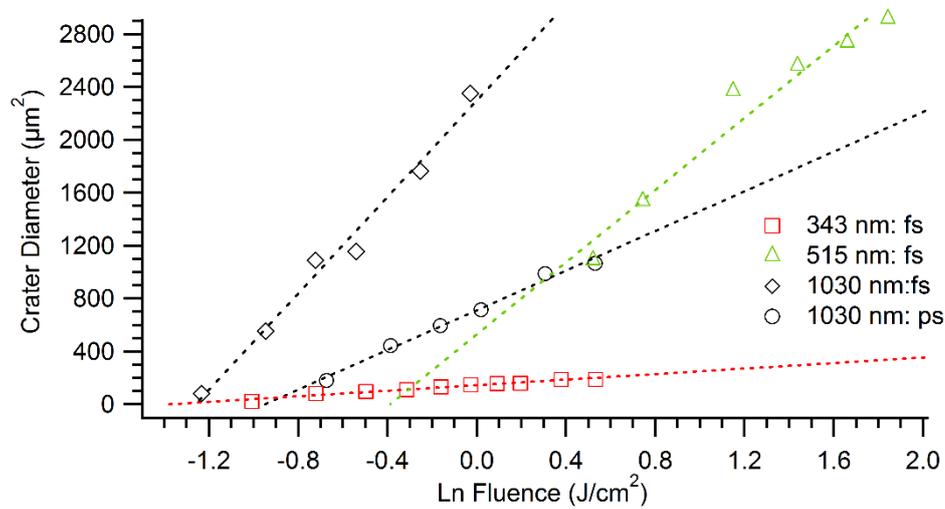

Figure 2: Fluence dependence of the crater diameter squared for ps and fs pulses, at wavelengths of 343, 515 and 1030 nm.

The threshold fluence and spot radii are calculated from this data using Liu's method [15]. The absorbed threshold fluence was calculated using equation (2). Table 3 presents the calculated values for the threshold fluence and absorbed threshold fluence.

Table 3: Relationship between the applied wavelength, threshold fluence, and absorbed threshold fluence.

| Wavelength (nm) | Threshold Fluence (J/cm²) | Absorbed Threshold Fluence $(mJcm^{-2})$ | Photon Energy (eV) | Film Absorption Coeff. $(m^{-1})$ | Spot Radius $\omega_0$ (μm) |
|---|---|---|---|---|---|
| 343 | 0.25 ± 0.02 | 12.5 ± 1 | 3.6 | 3.2×10⁶ | 10.2 ± 0.5 |
| 515 | 0.57 ± 0.04 | 9.68 ± 0.68 | 2.4 | 9×10⁵ | 24.5 ± 0.8 |
| 1030 (fs) | 0.32 ± 0.03 | 7.50 ± 0.70 | 1.2 | 1.2×10⁶ | 18.1 ± 2.3 |
| 1030 (ps) | 0.43 ± 0.02 | 10.1 ± 0.46 | 1.2 | 1.2×10⁶ | 23.1 ± 1.1 |

The threshold fluence shows a number of trends, in terms of the wavelength and pulse duration. At femtosecond pulse durations, the lowest threshold fluence is observed at 343 nm, with the highest observed at 515 nm. The threshold fluence also increases as the pulse duration is increased in the picosecond time regime. The absorbed threshold fluence, which takes into account the component absorbed in the film itself, shows a much different trend. For femtosecond pulses, the lowest threshold is now observed at 1030 nm, with the highest threshold at 343 nm.

## 3.2 ITO crater analysis after ultrashort laser irradiation

The surface topography of the craters was next investigated. Figure 3 shows the typical surface topography at wavelengths of 343, 515 and 1030 nm, in fs time regime, and at 1030 nm in the ps time regime. The applied fluence in each case was chosen to illustrate the cleanest removal of the film from the substrate, which in all cases required fluences above the calculated threshold fluence.

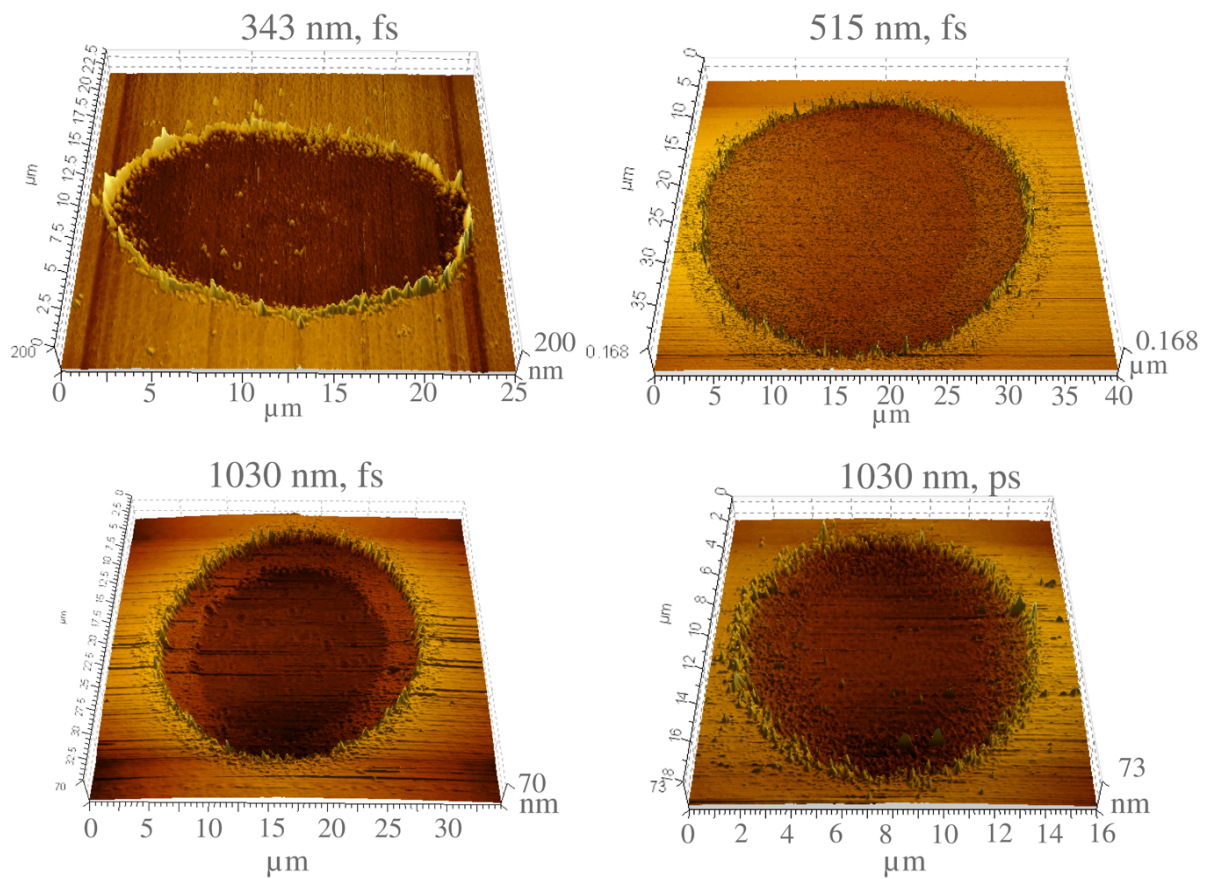

Figure 3 AFM surface topography of ITO on glass craters, showing selective removal of the ITO film, at wavelength of 343 (0.8 J/cm², fs), 515 nm, (1.4 J/cm², fs), 1030 nm (1.2 J/cm², fs) and 1030 nm (1.1 J/cm², ps).

In general, the craters were somewhat similar across the range of wavelengths and pulse durations investigated. In all cases, re-deposited material was noted in the crater area, in three cases the film appeared to be selectively removed from the substrate, except for the 1030 nm femtosecond pulse, where 5 – 10 nm of glass damage was noted. A number of different structures can be observed depending on the applied fluence, wavelength and pulse duration. Firstly, a particle-based ridge is observed at the crater edge at these applied fluences, which has height of 40 - 70 nm; the crater ridge was however shown to be dependent on the applied fluence. Secondly, in the femtosecond time regime for both 515 and 1030 nm wavelengths, two distinct regions were visible in the ablation area; an inner region, where the ITO film appears to be selectively removed, and an outer annulus, where incomplete removal of the film was observed. At the threshold fluence itself, the crater consisted almost exclusively of incomplete removal of the ITO film. The outer region where incomplete film removal occurs, had a nanostructured appearance with an increased surface roughness. The average depth of the structures is ≈ 5 nm. Randomly distributed spherical shaped nanostructures were also visible in the crater.

Further analysis of the crater edge and crater floor is shown in Figure 4, for 343 and 1030 nm fs pulses, and 1030 nm ps pulses. The applied fluence in each case is the same as in Figure 3.

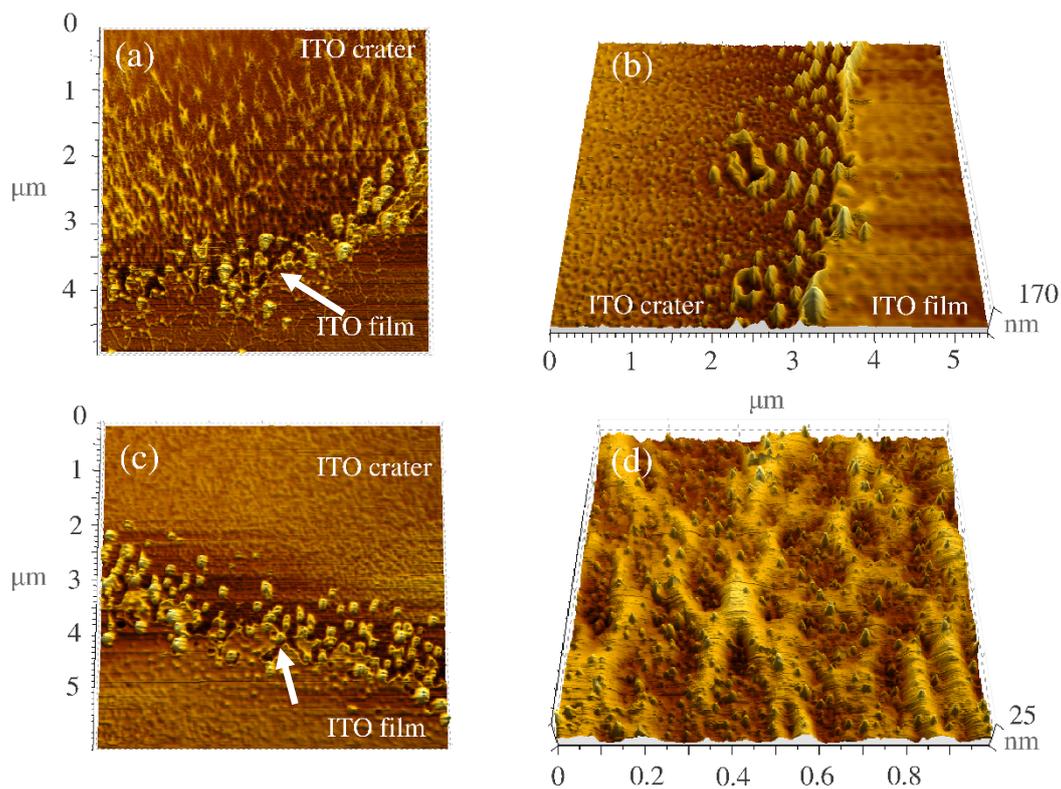

Figure 4 Crater edges for ITO on glass when ablated using (a) 1030 nm ps pulses, (b) 343 nm fs pulses (c) 1030 nm fs pulses. Large particles at the edge are visible at all wavelengths, including evidence of edge cracking at IR wavelengths (See arrows). Re-deposited nanoparticles on in center of crater floor (d) after 1030 nm ps pulses.

Analysis of the crater edge showed two main features. Evidence of a cracked crater edge was observed at the 1030 nm wavelengths. Also, large particles were observed at the crater edge, with sizes in the region of 100 - 200 nm. These larger particles are only observed at the crater edge, as illustrated in Figure 4 (b) and (c), with the majority deposited in the crater, as opposed to outside the crater. Smaller re-deposited nanoparticles are observed in the centre of the crater floor after laser irradiation at all wavelengths, as shown in Figure 4 (d) for 1030 nm ps pulses. The average diameter for these small nanoparticles on the crater floor was measured to be 20.4 ± 1.2, 19.2 ± 0.4, 29.1 ± 1.2 and 54.1 ± 2.2, for wavelengths at 1030 nm (ps), 1030 nm (fs), 515 nm (fs) and 343 nm (fs).

The surface topography of ITO on glass craters showed a high dependence on the applied fluence for both femtosecond and picosecond regimes. This is illustrated using 1030 nm ps pulses, in the region from the threshold fluence to 1.9 J/cm², as presented in Figure 5.

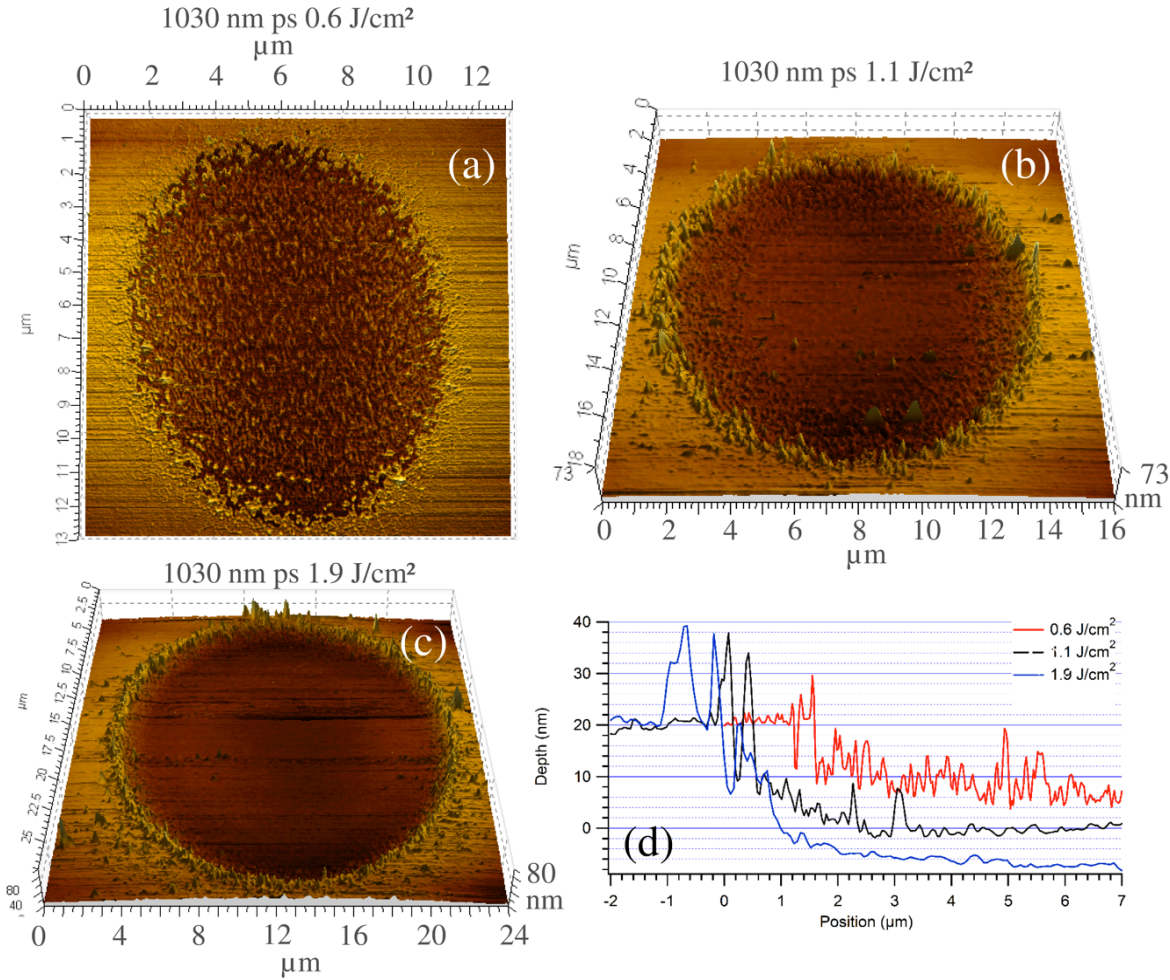

Figure 5: ITO on glass craters after ps laser irradiation with applied fluences of (a) 0.6 Jcm$^{-2}$ (b) 1.1 Jcm$^{-2}$ and (c) 1.9 Jcm$^{-2}$ (d) Cross sectional craters profile at applied fluences of 0.6, 1.1 and 1.9 Jcm$^{-2}$

At the lowest applied fluence of 0.6 J/cm$^{-2}$, in-complete removal of the film from the substrate was observed, with only 10 nm of the 20 nm ITO film being removed. The resulting crater floor was very rough, with a Ra value of 2.6 nm, with re-deposited nanoparticles with an average diameter of 20.4 ± 1.2 nm. The crater edge shows significant cracking, similar to the structures observed in Figure 4 (a). At an applied fluence of 1.1 Jcm$^{-2}$, complete selective removal of the film took place in the centre of the crater; the edge still showed evidence of cracking and appreciable roughness which was reduced towards the centre of the feature. At the highest applied fluence of 1.9 Jcm$^{-2}$, the crater edge showed less evidence of cracking; the edge consisted primarily of distinctive particles with no melt ridges. Approximately 10 nm of glass damage to the substrate was visible across the crater, which accounts for the smoother surface removal, with crater floor Ra roughness of 0.51 nm.

### 3.3 ITO on glass overlapped craters

Finally the effects of multiple overlapped pulses were examined for 1030 nm fs laser pulses. The applied fluence in this case was 1.2 Jcm$^{-2}$, as this was the optimal value suggested in from Figure 5. Figure 6 shows the surface topography of the ablated line where 3, 10 and 20 shots per laser spot area (SPA) were used. The corresponding depth profile along the scanned line is also shown in each case.

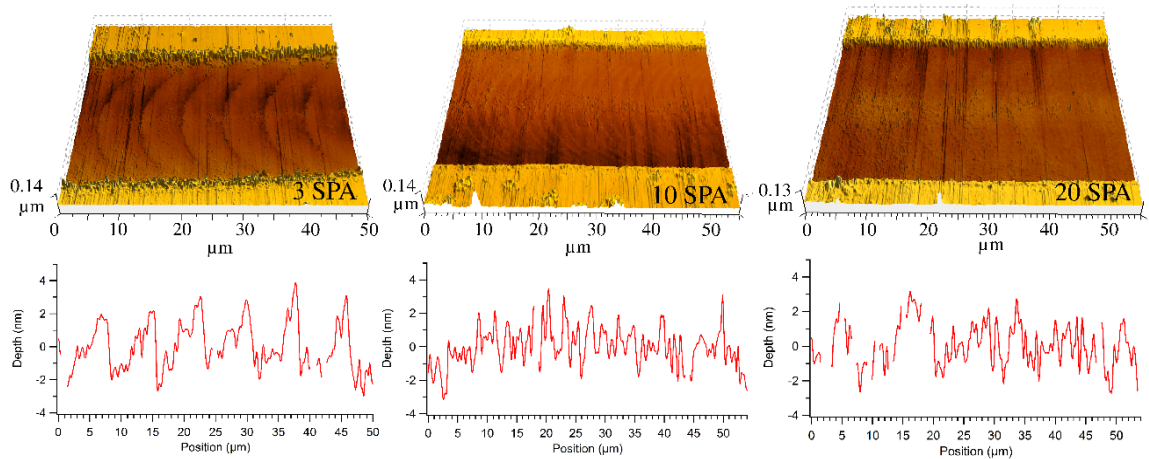

Figure 4 Femtosecond ITO on glass craters at 3, 10, 20 shot per area, at an applied fluence of 1.2 Jcm$^{-2}$. Line profiles present indications of surface topographies in centre along the direction of the scribe.

In general, clean removal of the film is seen from 3 SPA, with only minor visible improvements to the irradiated line at 20 SPA, in the form of a decreased edge height, and minor improvements to the crater floor consistency along the line. The surface roughness along the lines has a peak to trough maximum value of 4 nm, at all applied fluences.

### 3.6 ITO on glass two temperature simulation

Using the simple two temperature model described in equation (3) – (8), the temperature of the electrons and lattice was simulated during fs and ps laser irradiation of ITO on glass. The electron and lattice temperature was observed until equilibrium between the two systems was attained. Figure 7 (a) shows the peak temperature of the electron and lattice subsystems at the picosecond and femtosecond threshold fluences. Figure 7 (b) shows the spatial distribution of the temperature in the film and substrate, for fs pulses, at a measured applied threshold fluence of 0.32 Jcm$^{-2}$.

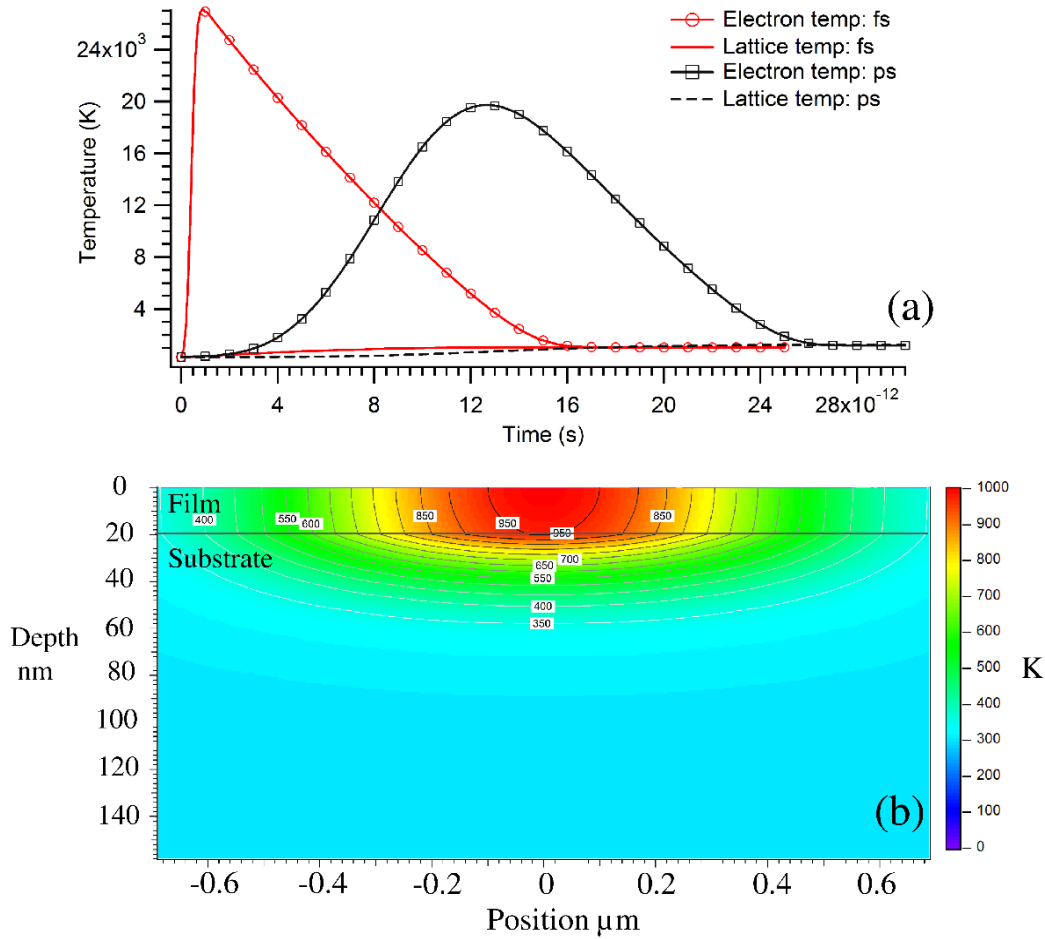

Figure 5 (a) Peak electron and lattice temperature after laser irradiation with fs and ps pulses, at their respective threshold fluences (b) Peak temperature in the film and substrate, 20 ps after irradiation with a 0.32 Jcm$^{-2}$ fs pulse.

The electron temperature was predicted to increase rapidly after the onset of laser irradiation. For the fs pulses, the peak temperature (27,000 K) of the electrons was predicted to occur (~1ps) when the lattice was still cold. For the ps pulses, the temperature of electrons was to found to heat more slowly, reaching a peak temperature of 19,600 K at ~13ps. The electronic heating and cooling for the ps pulses was almost Gaussian in nature, in contrast to the almost linear immediate decrease in temperature for fs laser irradiation. The time taken for equilibrium was found to depend on the pulse duration; for the fs and ps the time for electron lattice equilibration was approximately 16 and 26 ps respectively. The peak temperature of the lattice for fs and ps pulses is 1050 and 1200 K, respectively, at their appropriate threshold fluences. In both cases, the temperature was well below the melting temperature of ITO ($T_m = 1900 - 2100\ K$) [9].

The spatial distribution of the lattice temperature is shown in Figure 7 (b), which confirms some heat conduction across the interface. The temperature rapidly decreased with depth into the glass substrate, reaching room temperature at 60 nm below the substrate interface. However, up to 10 nm from the film substrate interface region, the temperature of the glass was predicted to be approximately equal to that of the film.

As the applied fluence was increased further, the peak temperature in the lattice increases. The predicted peak temperature in the lattice was examined for picosecond pulses across a range of applied fluences from 0.42 – 1.9 Jcm$^{-2}$, as presented in Figure 8.

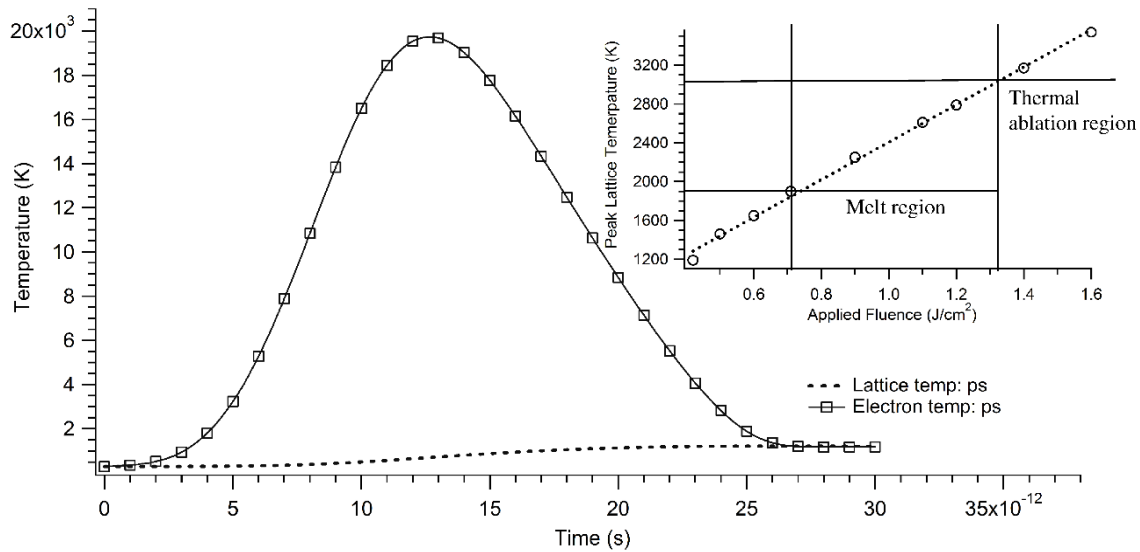

Figure 8 Electron and lattice temperatures after onset of laser irradiation with a 10 ps 1030 nm pulse, with an applied fluence of 1.9 Jcm$^2$. The peak lattice temperature is 4000 K Inset: Change in peak lattice temperature with applied fluence, from 0.42- 1.6 Jcm$^2$

The peak lattice temperature increases almost linearly in the region examined. A peak temperature of 1200 K is observed at the threshold fluence 0.43 Jcm$^{-2}$. The temperature does not reach the ITO melting temperature until an applied fluence of 0.71 Jcm$^{-2}$, defined as the beginning of the melting region in Figure 8. This melt region extends up to an applied fluence of 1.32 Jcm$^{-2}$, where the estimated ablation by thermal vaporisation region is predicted to begin.

**4. Discussion**

The study on ITO thin film patterning using femtosecond laser sources at wavelengths of 343, 515 and 1030 nm, at femtosecond and 1030 nm picosecond pulse durations, showed a dependence on laser processing parameters such as wavelength and pulse duration (Figure 2, Table 3). The lowest observed threshold fluence was at 343 nm, followed by 515 nm and then 1030 nm. The threshold fluence trend in this case follows the calculated absorption shown in Table 3. As expected, the highest absorption coefficient at 343 nm results in the lowest threshold fluence. The second highest absorption coefficient at 1030 nm results in a higher threshold fluence compared to 343 nm. This trend follows on to the 515 nm pulses. In the case of the 1030 nm picosecond pulses, the threshold fluence increases compared to 1030 nm femtosecond pulses by a factor of 1.34. This increase in threshold fluence with increasing pulse duration is typically observed for materials, and is due to the increase in heat diffusion length with increasing pulse duration [21]. This will result in greater heat loss to the glass substrate during absorption of the picosecond pulse, which will in turn result in an increase the threshold fluence. Taking into the account absorption coefficient and film reflectivity, the absorbed threshold fluence shows the opposite result, with the lowest absorbed fluence threshold occurring at 1030 nm. These values for absorbed fluence are an order of magnitude smaller than the corresponding values obtained for nanosecond pulses on the same films; 70.5 ± 3.4, 68.7 ± 5.2 and 50.2 ± 2.6 at wavelengths of 355, 532 and 1064 nm, respectively[1]. In order to understand this change in absorbed threshold fluence with wavelength, the absorption properties of the film must be considered. Absorption in the ITO thin film at all laser wavelengths is dominated by the high density of free carrier electrons in the conduction band. None of the laser wavelengths used in this study can excite band to band transitions in the thin film. This typically gives a free carrier absorption coefficient from the Drude model which is proportional to the wavelength squared $\alpha \sim \lambda^2$. Therefore, lower energy photons are more efficiently absorbed than higher energy photons. This agrees with the absorbed threshold fluence values observed in this study, where the lowest photon energy of 1.2 eV (1030 nm) results in the lowest absorbed threshold fluence, compared to 2.4 eV (515 nm) and 3.5 eV (343 nm).

In terms of the surface topography, a number of key features were observed (Figures 3 - 5). In general, the craters showed a near homogenous appearance over the wavelength range tested. In-complete removal of the 20 nm thick film was observed at applied fluences less than 0.7 Jcm$^{-2}$, for 1030 nm fs and ps pulses, with a total depth of 10 nm. This was accompanied by a fractured crater edge and re-deposition of nanoparticles across the crater floor; nanoparticles had a narrow distribution of diameters approximately equal to the grain size of the ITO film. Also, the re-deposited particles are angular in nature, with no evidence of spherical droplets. At mid-ranges of applied fluences (1 - 1.4 Jcm$^{-2}$), selective removal of the film was observed in the center portion of the crater, with in-complete film removal in the outer edge regions. Fractured crater edges were still observed, with no re-solidified melt ridges observable. At higher fluences (> 1.4 Jcm$^{-2}$), damage to the glass substrate was observed, typically in the region of 10 nm. The damage to the substrate results in what could be considered the smoothest removal of the film from the substrate, however, this is considered non-selective removal of the film, and is likely to be considered a sub-optimal process. Overall in terms of surface topography, no evidence of any thermal ablation processes was observed.

Reasonably clean removal of the ITO film was attained for overlapped pulses at 3 SPA, with only minor visible improvements to the irradiated line at 20 SPA (Figure 6); the height of each crater edge decreased slightly along the laser scanned line. The surface roughness along the lines had a peak to trough maximum value of 4 nm over a broad range of applied fluences investigated. These results are in stark contrast to those attained for selective patterning by nanosecond laser, where the requirement is for high shots per area to be used for reduced damage to the glass substrate[1].

The TTM model used in this study was used to simulate the rise in the electron and lattice temperatures for a number of parameters, such as applied fluence and pulse duration. Concentrating on the results for picosecond pulses (Figure 8), the TTM model predicts three regimes for film removal depending on the applied fluence. The simulated peak temperature in the film does not reach the ITO melting temperature for fluences up to 0.71 J/cm$_2$. At fluences higher than this, the peak temperature of the film will be greater than ITO melting temperature, leading to expected rapid melting of the film. This type of mechanism at the higher fluences would be expected to lead to evidence of re-solidified melt ridges after material cooling, however, this type of surface feature is not observed experimentally for these pulse durations. Likewise as the film removal mechanism might evolve from non-thermal to melting, and from melting to vaporization or phase explosion, which has been noted during high intensity ablation of dielectrics [22] and high fluence ablation of silicon [23]. If this is the case, then three distinct crater topographies should be observed, however this is not the case experimentally (Figure 3-5). No re-solidified crater ridges or re-deposited vaporised droplets are observed. Therefore we hypothesis that film removal is exclusively non-thermal in nature for these pulse durations. The ejection of material taking place on an ultrafast time scale before film phase change can take place.

As the applied fluence increases, the simulated peak temperature in the lattice becomes greater than the ITO melting and vaporisation points, which may result in the film removal via potential thermal ablation methods. Examining the TTM simulation, a thermal melt region is expected for applied fluences in the range of 0.71-1.32 J/cm$^2$. However, experimentally, no re-solidified melt ridge is observed at the crater edge. This indicates that ejection of the material may take place via the electron blast force stress before the melt can initiate. This leads to three competing process in the film after laser irradiation, the stress generated by the electron blast force, thermal heating of the film, and thermal conduction of heat across the substrate leading to glass damage. However, as no thermal melt processes are observed, ultrafast lattice deformation and grain ejection must take place before the film can heat to the ITO melting temperature, resulting in the ejection of the ITO grains before phase change can take place. The lack of melt ridges and ejected melt indicates that the traditional thermal ablation process may not play a role in film removal, leading to an exclusively stress generated removal mechanism. As the film is so thin in this case, at 20 nm, grain ejection of the full film can take place before the film has time to reach the melting temperature. Further studies are needed to observe how the film removal method changes with increasing film thickness. It is likely that film removal via ultrafast lattice deformation is only observed for very thin films, as seen in this study. If the film thickness was increased, there will most likely be a thickness were the all the grains cannot be ejected before the film reaches the melting temperature. This will lead to more traditional thermal film removal methods becoming more prominent.

**Conclusions**

An investigation of the ultra-short pulse laser ablation properties for 20 nm ITO thin films on glass substrates was undertaken. The threshold fluence for laser ablation was found to be highly dependent on the applied wavelength and pulse duration. The lowest absorbed threshold fluence was observed at 1030 nm, where more efficient coupling of the laser energy to the free electron carriers takes place. Film removal was primarily found to be through a non-thermal lattice deformation mechanism, which leads to the elastic break-up of the film at the ITO grain boundaries on a ps time scale. Film removal via melting and ablative methods is predicted by the TTM at higher fluences, however experimentally is not observed, indicating film grain breakup must take place before phase change in the film can take place. This non-thermal film ejection method results in the advantage of low SPA line removal of the film. This is in contrast to nanosecond laser pulses, which result in film removal via melting and vaporisation, which require high SPA overlaps to obtain clean line removal.

## Acknowledgements

This work is supported under an IRCHSS research grant no. PS/2010/2331.This work was conducted under the framework of the INSPIRE programme, funded by the Irish Government's Programme for Research in Third Level Institutions, Cycle 4, National Development Plan 2007-2013. Marie Curie Industry-Academia Partnership and Pathways (IAPP) project, Laser-Connect, funded under FP7-People-IAPP-2009-251542 and SFI Industry Research Fellowship RSF1319.